\documentclass[runningheads]{llncs}

\usepackage[namelimits]{amsmath} 
\usepackage{amssymb}             
\usepackage{amsfonts}            
\usepackage{mathrsfs}            

\usepackage[T1]{fontenc}
\def\doi#1{\href{https://doi.org/\detokenize{#1}}{\url{https://doi.org/\detokenize{#1}}}}
\usepackage{graphicx}
\usepackage[misc]{ifsym} 
\usepackage{listings}
\begin{document}
\title{FFCNet: Fourier Transform-Based Frequency Learning and Complex Convolutional Network for Colon Disease Classification}
\titlerunning{FFCNet.}
%
\author{Kai-Ni Wang \inst{1}, Yuting He\inst{2}, Shuaishuai Zhuang\inst{3}, Juzheng Miao\inst{1}, Xiaopu He\inst{3}, Ping Zhou\inst{1}, Guanyu Yang\inst{2,4}, Guang-Quan Zhou\inst{1} $^{(\textrm{\Letter})}$, Shuo Li\inst{5}}
%
\authorrunning{K. Wang et al.}
%
\institute{School of Biological Science and Medical Engineering, Southeast University, Nanjing, China\\
\email{guangquan.zhou@seu.edu.cn}\\
\and LIST, Key Laboratory of Computer Network and Information Integration (Southeast University), Ministry of Education, Nanjing, China
\and The first affiliated hospital of Nanjing medical university, Nanjing, China
\and Centre de Recherche en Information Biomédicale Sino-Français (CRIBs), Strasbourg, France
\and Department of Medical Biophysics, University of Western Ontario, London \\
ON, Canada
}
%
\maketitle              
\begin{abstract}
Reliable automatic classification of colonoscopy images is of great significance in assessing the stage of colonic lesions and formulating appropriate treatment plans. However, it is challenging due to uneven brightness, location variability, inter-class similarity, and intra-class dissimilarity, affecting the classification accuracy. To address the above issues, we propose a \textbf{F}ourier-based \textbf{F}requency \textbf{C}omplex \textbf{N}etwork (FFCNet) for colon disease classification in this study. Specifically, FFCNet is a novel complex network that enables the combination of
complex convolutional networks with frequency learning to overcome the loss of phase information caused by real convolution operations. Also, our Fourier transform transfers the average brightness of an image to a point in the spectrum (the DC component), alleviating the effects of uneven brightness by decoupling image content and brightness. Moreover, the image patch scrambling module in FFCNet generates random local spectral blocks,  empowering the network to learn long-range and local disease-specific features and improving the discriminative ability of hard samples. We evaluated the proposed FFCNet on an in-house dataset with 2568 colonoscopy images, showing our method achieves high performance outperforming previous state-of-the-art methods with an accuracy of $86.35\%$ and an accuracy of $4.46\%$ higher than the backbone. The project page with code is available at https://github.com/soleilssss/FFCNet.

\end{abstract}

\section{Introduction}
Accurate classification of early colon lesions plays an important role in diagnosis and treatment \cite{Zhang2016}. Colorectal cancer is usually diagnosed at an advanced stage due to insignificant early clinical symptoms, resulting in a high mortality rate \cite{Ladabaum2020,Elbediwy2016}. In clinical practice, colonoscopy is the most commonly used method to diagnose colorectal lesions \cite{Bibbins2016,Rex2016}. However, manual lesion classification is generally time-consuming and potentially subjective. Therefore, automated classification of colorectal lesions from colonoscopy images is critical in clinical analysis because it: \textbf{1)} helps physicians determine the type of colonic disease; \textbf{2)} formulates the most appropriate treatment options; \textbf{3)} compresses the duration of colonoscopy \cite{Marmol2017}. 

\begin{figure}
	\includegraphics[width=\textwidth]{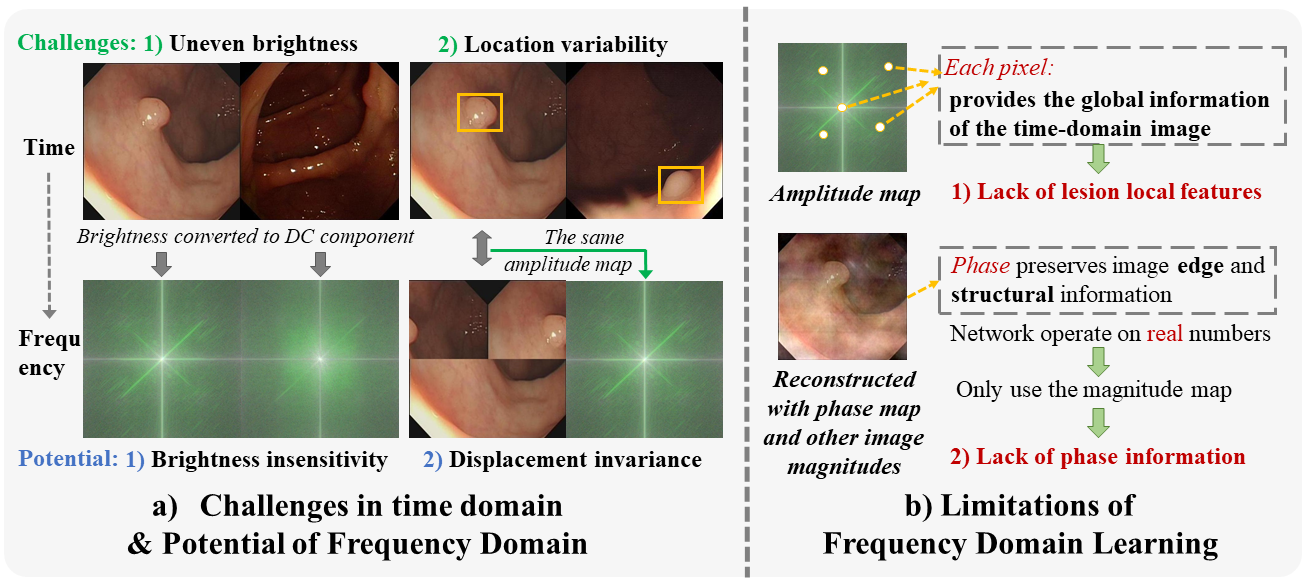}
	\caption{Frequency-domain learning has potential and limitations in colon disease classification tasks. a) Potential: Spectrum is brightness-insensitive and displacement-invariant. b) Limitations: Lack of image local information and phase information.} 
	\label{fig1}
\end{figure}

Existing research has achieved progress in classifying colon diseases \cite{Zhang2018,Carneiro2020}, but brightness imbalance and location variability remain intractable challenges (Fig.\ref{fig1} a)). The unbalanced illumination from the endoscope probe induces apparent color and brightness differences even in normal images, degrading the classification performance and increasing the difficulty in the model generalization. Some studies rely on data augmentation to resolve brightness imbalances \cite{Qadir2021}. Wei et al. designed the color exchange operation to force the model to focus more on the target shape and structure by generating images of various colors \cite{Wei2021}. However, the significant differences in brightness and color of endoscopic images prevent data augmentation from encompassing all distributions, resulting in poor performance in some distinct regions.  On the other hand, the locational variability of intestinal lesions is reflected in their appearance in various regions of the lumen wall. As a result, it is difficult for the network to learn all positional changes on small datasets and is prone to overfitting \cite{Liu2021}. Several works applied transfer learning to alleviate the location variability issues by acquiring features from large datasets \cite{Misawa2018,Wang2021}. Nonetheless, different feature distributions between source and target domain in transfer learning can lead to insufficient adaptability in network training.

Frequency learning has great potential for its brightness insensitivity and displacement invariance, which is recognized to improve the ability to discriminate the colon diseases (Fig.\ref{fig1} a)) \cite{Stuchi2020,Xu2020}. As we know, the DC component (only one value) in the spectrum represents the average brightness of the image \cite{Chi2020}. This breaks the correlation between image content and brightness, prompting the model to focus more on the target shape and structure. The magnitude map has displacement invariance and remains unchanged with the spatial movement of the time-domain image, which avoids the influence of changes in the lesion location. Moreover, the phase map provides contour and structural information of the image. \textbf{However}, the spectrum obtained by the direct Fourier transform of the image only contains global information, which limits the model to learn the local information of the lesion. Besides, the lack of phase information is caused by the fact that the network only learns from the magnitude spectrum due to the real number operation (Fig.\ref{fig1} b)).

In this study, we propose a novel frequency learning framework (FFCNet) for colon disease classification. Our work has the following contributions: \textbf{1)} Our method is one of the first to study the automatic classification of colon diseases (normal, polyps, adenomas, cancers) in the whole process. This four-level classification helps colonoscopists to determine the type of lesions accurately and advance the clinical diagnosis of early colorectal cancer. \textbf{2)} For the first time, we propose a framework that can be trained directly in the frequency domain by combining complex convolutional networks and frequency learning. The convolution kernels, blocks, and architectures in our complex network are modified into complex operations to enable direct learning of the spectrum with complex numbers, thus avoiding the loss of phase features caused by real network operations. Moreover, the spectral brightness insensitivity and displacement invariance have the ability to resolve the uneven brightness and positional variability of time-domain images. \textbf{3)} We innovatively present an image patch scrambling module embedded in FFCNet to generate local spectrograms. Spectral blocks provide local features of lesions so that the model has stronger discriminative ability of inter-class similarity. Also, through random shuffling operations, spectral patches that appear at different locations in the image will reveal long-range information to the model. \textbf{4)} The proposed method is competitive against well-known CNN architectures in experiments. This work has also sparked discussions on how classical CNN architectures can exploit spatial and frequency features in solving real-world problems to improve performance.

\section{Methodology}
\begin{figure}
	\includegraphics[width=\textwidth]{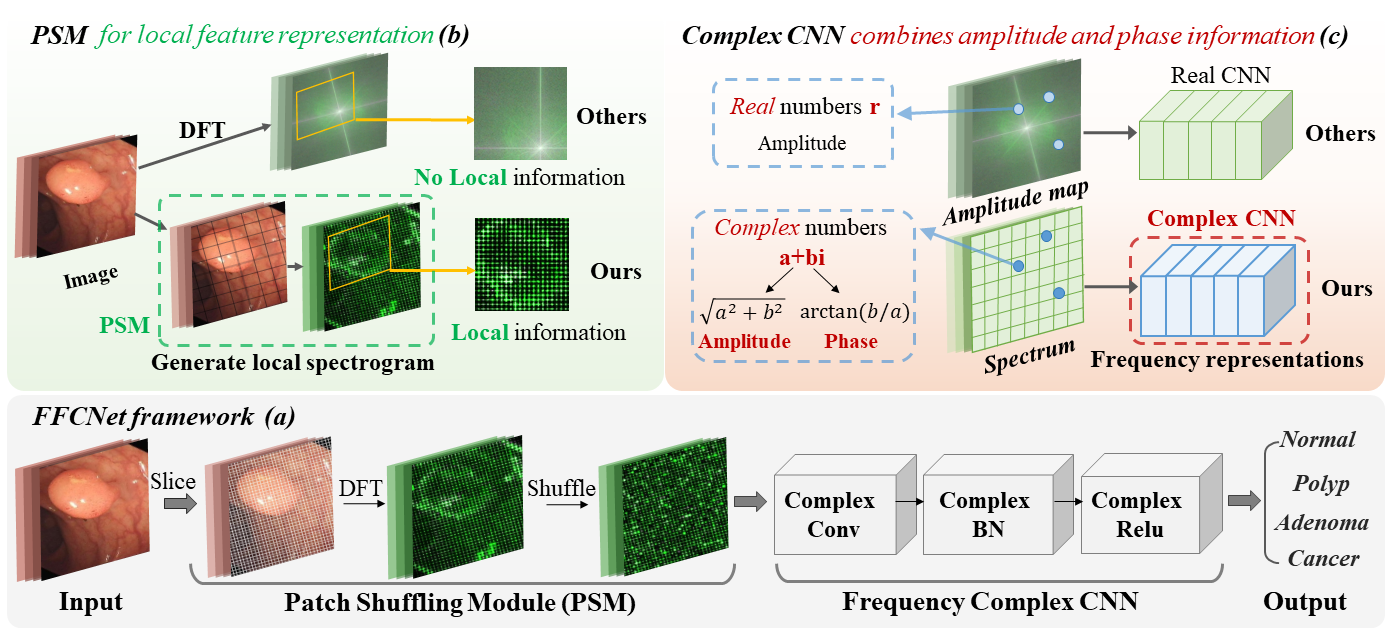}
	\caption{FFCNet is an end-to-end architecture consisting of a patch scrambling algorithm and a frequency complex CNN. (a) An overview of the proposed method. (b) The patch scrambling module which is introduced in Sect.\ref{2.1}. (c) The frequency-domain complex CNN which is introduced in Sect.\ref{2.2}.} 
	\label{fig2}
\end{figure}

FFCNet (Fig.\ref{fig2}) is composed of a patch scrambling module and a frequency-domain complex network. The patch scrambling module (Sect.\ref{2.1}) obtains a complex spectrogram by slicing the time-domain image after the Discrete Fourier Transform (DFT) and then scrambling, which effectively aggregates local information and improves the learning ability of non-local information. The frequency-domain complex network (Sect.\ref{2.2}) is capable of handling complex numbers based on the original network architecture. Specifically, replacing convolution, ReLU, batch normalization (BN) with complex convolution, complex ReLU, and complex BN enables the network to calculate complex spectrum to extract richer feature information. 

\subsection{Patch Shuffling Module (PSM)}\label{2.1}
The proposed PSM transforming time-domain images into frequency-domain representations consists of image dicing, DFT, and random shuffling. Compared with the image spectrum without dicing, the network will only learn global information due to the spectral characteristics that a point in the frequency domain can affect the entire image, which neglects the necessary local features for colon classification. Hence, as shown in Fig.\ref{fig2} b), we perform DFT after slicing images to guide the network to focus on recognizable local features. On the other hand, the scrambled spectrum block further improves the long-distance feature learning of the model.

Given an input image I, we first uniformly partition the image into $K \times K$ patches denoted by matrix R. $R_{ij}$ denotes an image patch where i and j are the horizontal and vertical indices, respectively ($1 \leq i \leq K$, $1 \leq j \leq K$). After the dicing, each block will be transformed to the frequency domain. Then, for each image block m, denoted as $f_{m}\left( x,~y \right)$, with size $M \times N$, the DFT is computed according to the following expression:
\begin{equation}
F_{m}\left( {u,v} \right) = {\sum\limits_{x = 0}^{M - 1}{\sum\limits_{y = 0}^{N - 1}{F_{m}\left( {x,y} \right)e^{- j2\pi{({\frac{ux}{M} + \frac{vy}{N}})}}}}}
\end{equation}
Finally, the spectral patch will be randomly shuffled with probability $p$. Since the neatly arranged spectrum has been corrupted, in order to identify these randomly arranged spectral blocks, the classification network has to find discriminative regions and identify small differences between classes.

\noindent \textbf{Summarized advantages:} Our PSM compensates for both local and long-range features of the image while preserving the advantages of the frequency domain. In addition, the local spectrogram has a smaller numerical distribution range than the original spectrogram, which improves the convergence degree of the gradient and speeds up the training of the model.

\subsection{Frequency-domain complex network}\label{2.2}

The proposed frequency learning network can directly learn complex-valued spectrograms through complex operations during training. The backbone architecture of the complex network adopts ResNet \cite{He2016}, and the internal operations are replaced by complex sub-components (complex convolution, ReLU, batch normalization). Each residual block consists of two 3*3 convolutional layers and one connection path. Thus, 
the proposed network takes into account the advantages of frequency information features and the rich expressive power of complex operations.

\noindent \textbf{Complex convolution.}
In order to perform an operation equivalent to the traditional real-valued 2D convolution in the complex domain, the real part $a$ and the imaginary part $b$ of the complex matrix $P = a + bi$ in the spectrogram are respectively input into the network. Meanwhile, two sets of convolution kernels $c$ and $d$ are inserted to simulate the real and imaginary parts of the complex convolution kernel $Q = c + di$. Complex convolution can be expressed as:
\begin{equation}
P*Q = \left( {a + bi} \right)*\left( {c + di} \right)\\
	= \underset{real}{\underbrace{\left({a*c - b*d} \right)}} + \underset{imaginary}{\underbrace{\left( a*d + b*c \right)}}i\\
\end{equation}
Where a,b,c,d are all real numbers.

\noindent \textbf{ Complex ReLU.}
The neural network relies on the ReLU function to introduce nonlinearity to promote the sparsity of the network. The ReLU function sets all negative values in the matrix to zero and does not change the remaining values. The complex ReLU is the addition of the real and imaginary parts after applying ReLU respectively. Complex ReLU satisfies the Cauchy-Riemann equation when both the real and imaginary parts are strictly positive or strictly negative \cite{Trabelsi2017}. The specific formula is as follows:
\begin{equation}
	ReLU\left( P \right) = \underset{real}{\underbrace{ReLU\left( a \right)}} + \underset{imaginary}{\underbrace{ReLU\left( b \right)}}i
\end{equation}

\noindent \textbf{Complex BN.}
BN is often employed to accelerate learning in neural networks. BN forcibly pulls the distribution of the input values of each layer of neural network back to a standard normal distribution with a mean of 0 and a variance of 1. For BN of complex numbers, it is unreasonable to translate and scale it so that it has a mean of 0 and a variance of 1. This normalization does not ensure that the variances of the real and imaginary parts are equal. It will be oval, possibly with high eccentricity. Hence, we treat it as a two-dimensional vector to change the data distribution. 

Given a batch input $x$ to compute the mean and variance, the normalized $\overset{\sim}{x}$ is expressed as:
\begin{equation}
	\overset{\sim}{x} = \frac{\left( x - E\left( x \right) \right)}{\sqrt{C}}
\end{equation}
Where $c$ is the covariance matrix and $E\left( x \right)$ is the mean of the data. $c$ is a $2 \times 2$ matrix represented as:
where $R\left( x \right)$ and $I\left( x \right)$ represent the real and imaginary parts of $I\left( x \right)$, respectively.
Similar to real number normalization, learnable reconstruction parameters $\gamma$ and $\beta$ are introduced to restore the feature distribution to be learned by the network. The difference is that the shift parameter is a complex parameter with two learnable components (real and imaginary). The scaling parameter is a $2 \times 2$  positive semi-definite matrix with only three degrees of freedom. There are three learnable components. The complex BN is defined as:
\begin{equation}
	BN\left( x \right) = \gamma\left( \overset{\sim}{x} \right) + \beta
\end{equation}	

\noindent \textbf{Summarized advantages:} Our elaborate complex CNN implements a full range of frequency-domain analysis, learning both magnitude and phase information in a unified architecture. Not only that, complex convolution, ReLU, and BN maintain the strength of easier optimization of complex numbers, and further improve the expressive ability of frequency features.

\section{Experiments and Results}
\textbf{Experiment protocol.}
\textbf{1)Datasets.} The study included 3568 standard white-light endoscopic images including 865 normal, 843 polyps, 896 adenomas, and 964 cancers. We randomly split the dataset into training, validation and testing in a 6:2:2 ratio. \textbf{2)Settings.} We use version 1.1 of PyTorch \cite{Paszke2019} to perform all experiments on NVIDIA TITAN X (PASCAL) GPU machines and evaluate our proposed method on the widely used classification backbone network ResNet-18. Input images are resized to a fixed size of 400×400. Random horizontal and vertical flips 
were applied for data augmentation. During training, we trained all network by SGD optimizers with a learning rate of 0.1 and a mini-batch size of 32 for 600 epochs. The probability p of patch shuffling was set to 0.3. During testing, data augmentation and patch scrambling algorithms were disabled. The diced spectrum of the original image was fed into the complex classification network for final prediction. \textbf{3)Evaluation metrics.} We evaluate the classification performance using four metrics: Accuracy, Precision, Recall
and F1-score. More details are in our \emph{Supplementary Material.}

\begin{table}
\caption{FFCNet yields higher performance than different classical classification methods on each metric (\%).}\label{tab1}
\centering
\setlength{\tabcolsep}{3.5mm}{
\begin{tabular}{lllll}
\hline
& Accuracy       & Precision      & Recall         & F1-score       \\
\hline
ResNet \cite{He2016} & 81.89          & 81.96          & 81.89          & 81.91          \\
MobileNet \cite{Howard2017} & 81.11 & 81.14 &81.11 & 81.07 \\
EfficientNet \cite{Tan2019} & 84.44 &84.76 &84.44 &84.55 \\
DenseNet \cite{Huang2017} & 84.86          & 84.86          & 84.86          & 84.80 \\
GoogLeNet \cite{Szegedy2017} & 85.42          & 85.72          & 85.42          & 85.52          \\ 
\hline
CoAtNet \cite{Dai2021} & 85.93	&86.08&	85.93&	85.94\\
Fast \cite{Dai2021} & 81.57 & 81.86 & 81.57 & 81.60\\
GFNet \cite{Rao2021} & 84.81 & 84.92 & 84.81 & 84.86\\
K-Space \cite{Han2019}& 83.28 & 83.42 & 83.28 &83.24\\
\hline
FFCNet & \textbf{86.35} & \textbf{86.61} & \textbf{86.35} & \textbf{86.44}\\
\hline
\end{tabular}}
\end{table}

\noindent \textbf{Comparative experiments show the superiority of our FFCNet:}
The comparison of our network with several classical classification methods shows great potential for application in colonoscopy classification scenarios. Compared with other methods (Table 1), FFCNet has the highest accuracy ($86.35\%$), precision ($86.61\%$), recall ($86.35\%$) and F1-score ($86.44\%$). In particular, the accuracy of FFCNet is $4.46\%$ higher than that of the backbone network ResNet, indicating that the frequency features provide a significant improvement to the architecture. 

\begin{figure}
	\includegraphics[width=\textwidth]{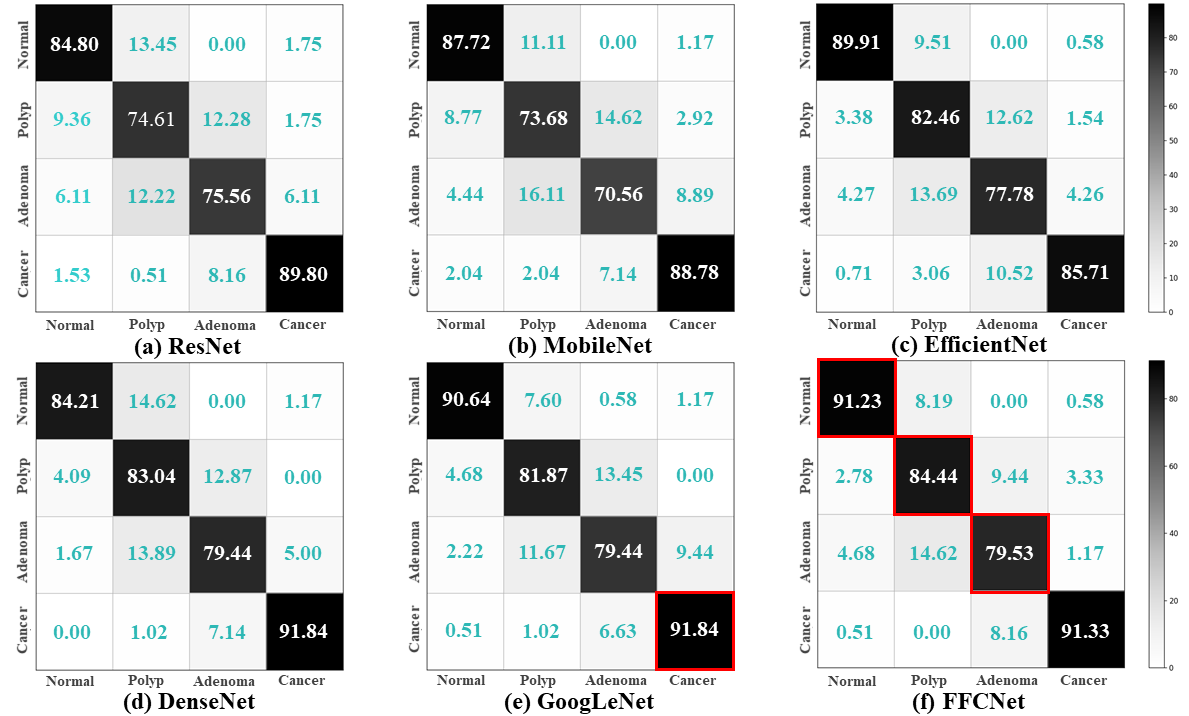}
	\caption{The results of the confusion matrix illustrate that FFCNet obtains excellent performance in all categories. The numbers in the confusion matrix represent the percentage of predicted classes.} 
	\label{fig3}
\end{figure}

Comparisons with frequency, complex and transform networks also indicate the superiority of our FFCNet: a joint CNN and transformer network (CoAtNet, $85.72\%$), a CNN network with added frequency modules (Fast, $81.57\%$), a transformer network with added frequency modules (GFNet, $84.81\%$), and other complex networks (K-Space, $83.28\%$). FFCNet outperforms them without requiring pre-training, indicating that the designed model is more efficient.

The results of the confusion matrix demonstrate the superiority of FFCNet in discriminating similarity between classes. Our network obtains excellent performance in all categories Fig.\ref{fig3}. Not only that, the network achieves the highest accuracy in the most difficult to distinguish polyps ($84.44\%$) and adenomas ($79.53\%$), respectively. This finding is attributed to our PSM guiding the network to acquire disease-specific information by learning local and long-range features.

\begin{figure}
	\includegraphics[width=\textwidth]{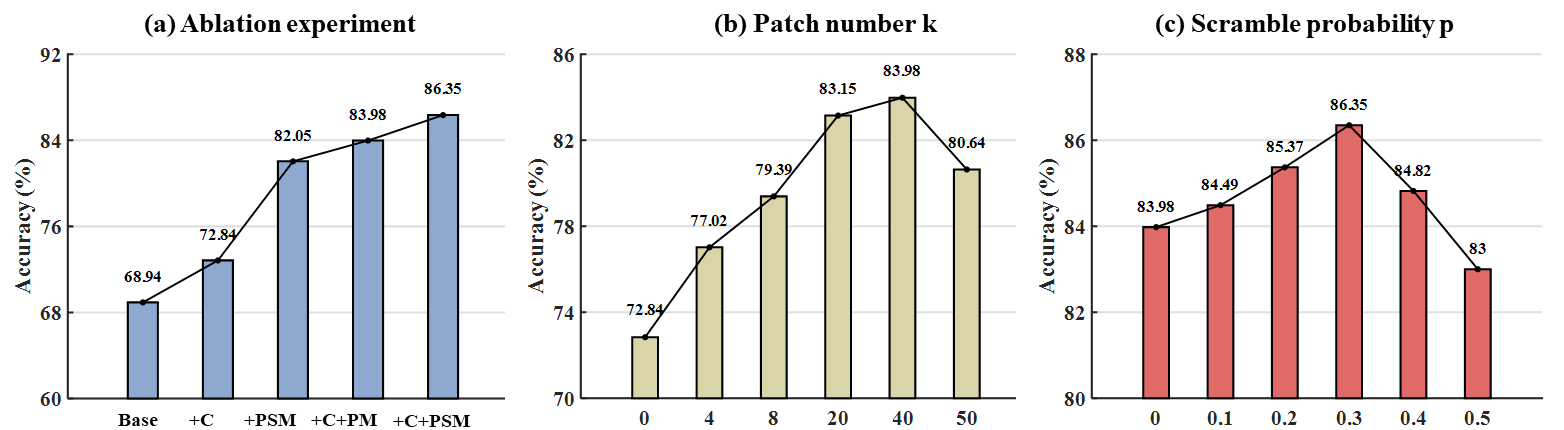}
	\caption{Ablation experiments and hyperparameter experiments on the test set illustrate the necessity of submodules and the influence of parameters in the network, respectively. In Figure (a), 'C' and 'PSM' represent the complex network and the patch scrambling module, respectively. 'PM' is the PSM removal scrambling.} 
	\label{fig4}
\end{figure}

\noindent\textbf{Ablation experiments demonstrate the contribution of the proposed module:} Fig.\ref{fig4} a) shows the average accuracy of the ablation experiments quantitatively, demonstrating that each submodule contributes to the performance improvement. We first build a baseline model using magnitude maps with ResNet and gradually incorporate each of the submodules discussed in Section 3 into the baseline model. After adding PSM alone, local information is supplemented resulting in an accuracy improvement of $9.2\%$ and $4.3\%$, compared to the baseline and baseline models with complex networks. Incorporating our proposed complex network to the baseline model improved by $3.9\%$ verifies the ability of complex networks to mine phase information. The last two columns validate the importance of random shuffling significantly improving classification accuracy by learning distance information.

\noindent\textbf{Hyperparameter experiments analyze the superiority of the architecture:} We evaluate the effect of patch number and random scramble probability on the network (Fig.\ref{fig4} b) c)). The slicing operation brings local information of the image, so the accuracy of the network gradually increases as the number of slices grows. However, if the image patch is too small, it will destroy the advantages of frequency domain features and reduce 
model performance. Likewise, random shuffling allows the network to learn from information over long distances, yet the difficulty of network learning also increases. Therefore, we shuffle the image to stabilize the network performance with a certain probability.

\section{Conclusion} In this study, we propose a novel method to advance colon disease classification in colonoscopy images from a frequency domain perspective. The proposed framework, FFCNet, introduces complex convolution operations, enabling the network to directly operate on complex spectra to obtain rich texture features and eliminate the influence of brightness imbalance. Furthermore, the patch scrambling algorithm we developed preprocesses the spectrogram so that the network can learn both long-range and local information. Finally, we compare the performance of the framework with other methods. The results show that our frequency-domain complex number framework is competitive with time-domain models in diagnosing colon diseases.

\subsubsection{Acknowledgements} This work was supported by the National Key R\&D Program Project (2018YFA0704102).

%
%
%
%

\end{document}